\documentclass[12pt]{iopart}

\usepackage[10pt]{type1ec}  
\usepackage[T1]{fontenc}
\usepackage{CJKutf8}
\usepackage{amssymb}
\usepackage[overlap, CJK]{ruby}
\usepackage{CJKulem}
\newenvironment{SChinese}{%
  \CJKfamily{gbsn}%
  \CJKtilde
  \CJKnospace}{}

\usepackage{color}
\usepackage{graphicx}

\newcommand{\ba}{\begin{array}}
\newcommand{\ea}{\end{array}}

\newcommand{\ptrans}{p_{\rm trans}}

\newcommand{\Mmax}{M_{\rm max}}

\newcommand{\Msolar}{{\rm M}_{\odot}}
\newcommand{\Mchirp}{{\mathcal M}}

\newcommand{\cQM}{{c^{\phantom{1}}_{\rm QM}}}

\newcommand{\cQMsq}{c^2_{\rm QM}}

\newcommand{\al}{\alpha}

\newcommand{\De}{\Delta}
\newcommand{\ep}{\varepsilon}
\newcommand{\eps}{\epsilon}

\newcommand{\La}{\Lambda}

\begin{document}

\title[J Phys G review on quark matter in neutron stars]{Signatures for quark matter from multi-messenger observations}

\author{Mark G. Alford}
\address{Physics Department, Washington University, St.~Louis, MO~63130, USA}
\author{Sophia Han 
(\begin{CJK}{UTF8}{}\begin{SChinese}韩 君\end{SChinese}\end{CJK})
}
\address{Department of Physics and Astronomy, Ohio University, Athens, OH~45701, USA}
\address{Department of Physics, University of California Berkeley, Berkeley, CA~94720, USA}
\author{Kai Schwenzer}
\address{Department of Astronomy and Space Sciences, Istanbul
  University, Beyaz\i t, 34119, Istanbul, Turkey}


\begin{abstract}
We review the prospects for detecting quark matter in neutron star cores.
We survey the proposed signatures and emphasize the importance of data from neutron star mergers, which provide access to dynamical properties that operate on short timescales that are not probed by other neutron star observables.
\end{abstract}

\maketitle

\section{Introduction}
\label{sec:intro}

The interior of compact stars is the most extreme environment in nature, characterized by the highest matter densities, huge spacetime curvature, and potentially gigantic magnetic fields. It is also one of the least understood regimes of matter because of our present inability to solve Quantum Chromodynamics (QCD), the theory describing the dominant interactions, in this regime \cite{Brambilla:2014jmp}. 

The weakening of the QCD coupling at high energies leads us to expect novel forms of deconfined matter at high density whose active constituents are quarks rather than baryons. However, because the coupling runs logarithmically slowly, QCD at the typical density of compact stars is still strongly coupled and theorists are unable to predict whether deconfined quark matter will exist in their interior. Astrophysical observations of compact stars are therefore crucial to determine the composition and properties of dense matter. Such insight into matter at high density could also help us to improve our limited understanding of strong interaction phenomena in ordinary matter such as confinement, spontaneous chiral symmetry breaking, and dynamical mass generation.  

\subsection{Distinguishing the phases of dense matter}
In the traditional Landau classification, different phases of matter are distinguished by their transformation properties under exact symmetries. The exact symmetries relevant to compact stars are: quark phase invariance leading to baryon number conservation (spontaneously broken in superfluids); electromagnetic gauge symmetry (spontaneously broken in superconductors); and spacetime translation and rotation symmetries (spontaneously broken in crystalline or anisotropic phases). In practice, however, there can be other significant differences between forms of matter. These include the breaking of approximate symmetries such as chiral symmetry, which is manifest in anomalously light pseudo-Goldstone bosons, and differences in the nature and interactions of low-energy degrees of freedom (examples in terrestrial condensed matter
include the Lifshitz \cite{Volovic:2017,Volovik:2006gt} or metal/insulator \cite{Wen:2005,Dobrosavljevic:2011} transitions) where the Fermi surface of unpaired particles, which is protected against small perturbations by a topological quantum number, similar to those arising from exact symmetries, can change its topology: e.g. disappear or become gapped. All these can result in observable signatures such as different power-law dependencies on small parameters (e.g.~$T/\mu$ or  mass ratios or couplings) that result in different behavior of material properties.

\begin{figure}
\includegraphics[width=0.6\hsize]{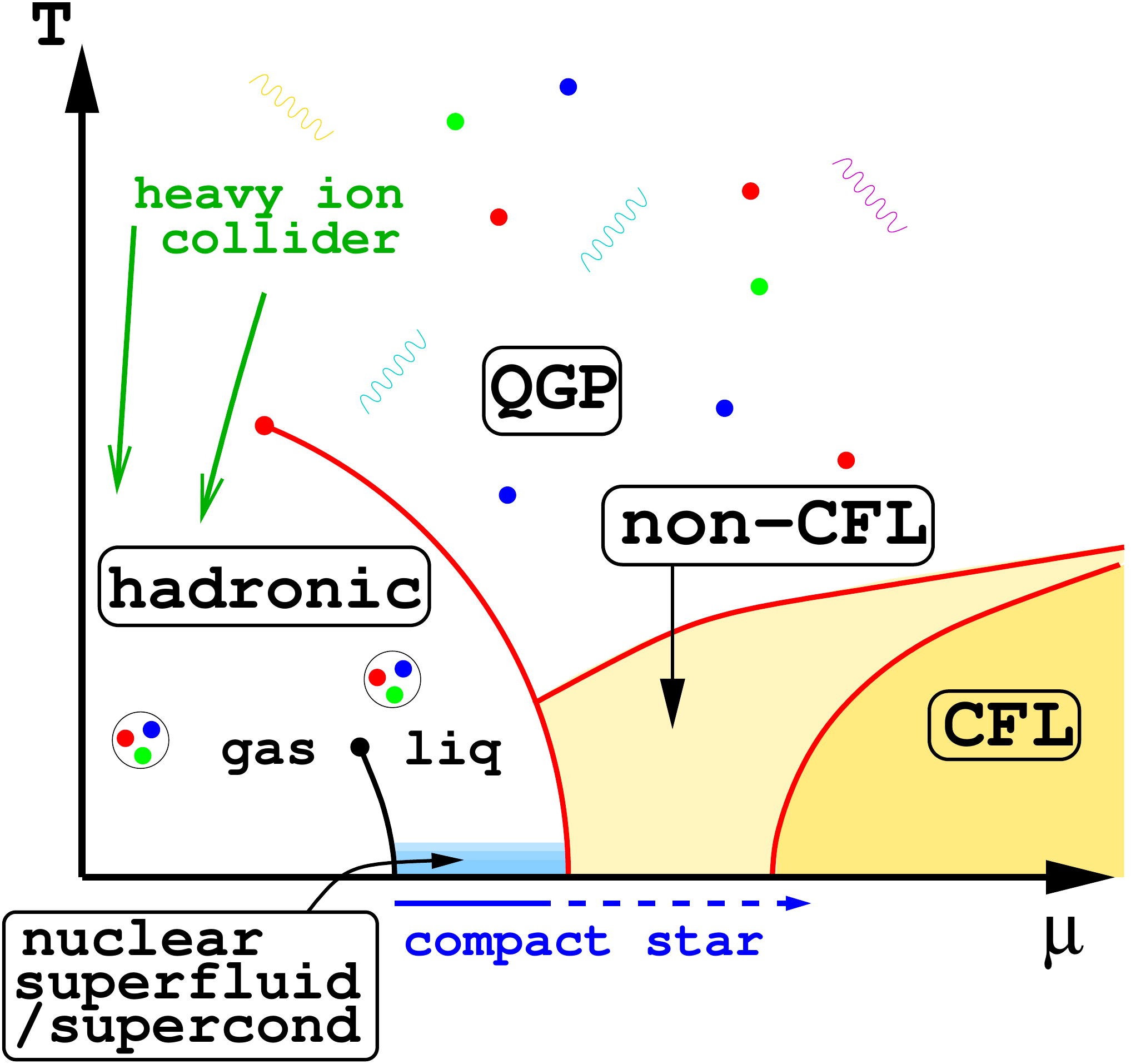}
\caption{Conjectured diagram of phases of matter as a function of
baryon chemical potential $\mu$ and temperature $T$.
At the highest densities, we expect the color-flavor-locked (CFL) phase of quark matter.
}
\label{fig:phase diagram}
\end{figure}

Fig.~\ref{fig:phase diagram} shows a conjectured outline for the phase diagram of matter as a function of quark chemical potential $\mu$ and temperature $T$. Compact stars probe the low-temperature part of the diagram where at quark chemical potential $\mu\approx 308\,{\rm MeV}$ there is a first-order transition from hadronic gas to hadronic liquid (nuclear matter), and at high chemical potential we expect phases of ``quark matter'' where quarks are no longer confined inside baryons and undergo Cooper pairing to form color superconducting \cite{Alford:2007xm} phases. At the highest densities we expect the color-flavor-locked phase \cite{Alford:1998mk}, in which up, down, and strange quarks form Cooper pairs in a particularly symmetric pattern.

Formally, ``quark matter'' is not a straightforwardly distinguishable phase of matter. In pure gauge theories there is an order parameter, the Polyakov loop, which signals deconfinement via the breaking of the center symmetry of the gauge group, but when quarks are present they screen the confining force and the Polyakov loop is always non-zero \cite{Brambilla:2014jmp}.

Color superconducting phases of quark matter are also not characterized as a class by any single order parameter. Color superconductivity gives mass to the gluons so color superconducting quark matter can be thought of as the ``Higgs'' phase of QCD, but there is no order parameter that distinguishes the confined phase from a Higgs phase whose condensate transforms as the fundamental representation of the gauge group \cite{Fradkin:1978dv}. However, certain color superconducting phases may have characteristic, if not unique, properties. In terms of the Landau classification based on exact symmetries and order parameters, both nuclear matter and quark matter have phases that are superfluid, superconducting, or crystalline.
In terms of approximate symmetries, the CFL quark matter phase spontaneously breaks chiral symmetry, but the two-flavor Cooper paired (``2SC'') phase does not \cite{Alford:2007xm}.
Examples of Lifshitz or quantum ordering are the appearance of muons at sufficiently high density, or the transition from unpaired quark matter to the 2SC phase.

Thus, although Fig.~\ref{fig:phase diagram} shows a first-order transition from nuclear matter to quark matter, there may equally well be a crossover region: a range of densities that is no more easily understood in terms of hadrons than in terms of quarks \cite{Yamamoto:2007ah,Abuki:2010jq}. 
For example, one could go continuously from a superfluid non-superconducting hadronic insulator (understood in terms of Cooper pairing of neutrons, protons, and hyperons) to the CFL phase of quark matter which is also a superfluid insulator \cite{Schafer:1998ef,Alford:1999pa}. We note that such a continuous crossover in the phase diagram is different from a mixed phase (Gibbs construction) which is a manifestation of a first-order transition in the phase diagram.

Considering all these diverse options, the detection of quark matter will require a comprehensive framework that enables us to find---if not qualitative, at least clear quantitative---differences that distinguish forms of quark matter from {\it all} possible forms of hadronic matter. The corresponding signatures should be robust enough to enable a definite discrimination, despite any uncertainties in the astrophysical data.
As an example of this sort of comprehensive classification, we show in Table~\ref{tab:shear-viscosity} a compilation of the shear viscosity in various possible phases of hadronic and quark matter. As we will discuss below, shear viscosity is relevant to astrophysical observables such as spin down. By combining many such calculations of microscopic properties to make predictions of the relevant astrophysical observables we hope to build a framework that can take into account all relevant uncertainties to formulate a well defined statistical test to 
convincingly reject the null hypothesis (that neutron stars consist entirely of hadronic matter) at a definite significance level.

\begin{table}
\begin{tabular}{lcll}
& \multicolumn{2}{c}{{\bf Shear Viscosity}} \\
Phase & Dominant processes & Shear viscosity & Ref \\
\hline
Nuclear: ungapped & $\ba{c} e\ e \to e\ e\\ N\ N \to N\ N \ea $ 
 & $\ba{rl}\eta \sim & (T/\mu)^{-5/3}\\ + & (T/\mu)^{-2} \ea$ & \cite{Shternin:2008es} \\[1ex]
Hyperonic: ungapped & $\ba{c} e\ e \to e\ e\\ N\ N \to N\ N \ea $ 
 & $\ba{rl}\eta \sim & (T/\mu)^{-5/3}\\ + & (T/\mu)^{-2} \ea$
 & \cite{Shternin:2008es} \\[1ex]
Nuclear: superfluid & $e\ e \to e\ e$ & $\eta \sim  (T/\mu)^{-5/3} $ & \cite{Shternin:2008es} \\[1ex]
Quark: ungapped & $q\ q \to q\ q$ &  $\eta \sim  (T/\mu)^{-5/3} $ & \cite{Heiselberg:1993cr} \\[1ex]
Quark: CFL  & $\phi\ \phi \to \phi\ \phi$ &  $\eta \sim  (T/\mu)^{4} $ & \cite{Manuel:2004iv} \\
\end{tabular}
\caption{Dependence of shear viscosity on temperature at low temperatures ($T\ll\mu$) in various phases of hadronic and quark matter.
\label{tab:shear-viscosity}}
\end{table}

\subsection{Astrophysical signatures of quark matter}
The microphysical characteristics that distinguish different forms of matter, as outlined above, will, we hope, lead to astrophysical signatures in the observable behavior of neutron stars that could allow us to infer which forms of matter are present in their interiors. In particular, dynamic material properties like dissipation, transport, or emission are sensitive to the low-energy degrees of freedom and their interactions, which can differ dramatically between the various forms of dense matter. 
We now give an overview of possible signatures, with details in the sections below.

\begin{enumerate}
\item Static observables based on the equation of state. \\
Static observables including mass, radius, moment of inertia, tidal deformability and density profile are completely determined by the equation of state which is not in general particularly sensitive to the phase of the material. However, if there is a first-order transition between phases of very different density then this could verifiably affect the mass-radius relation. 
\item Thermal evolution. \\
The cooling history of stars tends to be dominated by the phase with the highest neutrino luminosity. This could allow us to identify classes of dense matter phases, but it will be hard to distinguish hadronic from quark matter exclusively from their cooling behavior,
since there are fast cooling mechanisms in both hadronic and quark matter (direct Urca neutrino emission in ultra-dense hadronic matter being the fastest known mechanism) and the phase space of such reactions can in both cases be arbitrarily reduced by momentum restrictions due to the Fermi seas of the involved particles or by pairing. 

\item Oscillations and spin-down.\\
Gravitational wave astronomy based on global oscillation modes, like the r-mode, is a particularly promising method to distinguish phases of dense matter since the gravitational waves probe the dense interior. Even without a direct gravitational wave detection, electromagnetic probes indirectly inform us about damping of oscillations in dense matter, which can vary drastically between standard neutron matter and exotic forms of matter. There are now strong indications that a minimal neutron star scenario cannot provide the strong dissipation required to explain the accumulated astrophysical data.

\item Solid phases.\\
Glitches are thought to be related to solid phases coexisting with superfluid, but the lack of an in-depth understanding makes it hard to use them as evidence of the crystalline phase of quark matter. A better signature would be continuous gravitational wave emission from the rotation of a star containing a core of crystalline quark matter, which is much more rigid than any other known hadronic phase and hence supports larger deformations.

\item Neutron star mergers.\\
We have now detected gravitational wave and electromagnetic emission from the inspiral phase of neutron star merger. Dynamic material properties of the dense matter vary greatly between hadronic and quark phases, and will greatly influence the evolution of the highly excited merger product and the future gravitational wave signal of the actual merger phase. 
\end{enumerate}

Taking into account that there is not a single, unique criterion, it will likely be necessary to combine the available data on the various astrophysical processes and observables discussed above to positively confirm the presence of quark matter in compact stars. 

\section{Static observables based on the equation of state}
\label{sec:EoS}

The global structure of static neutron stars, including properties such as the mass, radius, tidal deformability, and the moment of inertia, are obtained for a given equation of state (EoS) by solving the equations of hydrostatic equilibrium, the Tolman-Oppenheimer-Volkov (TOV) equations \cite{Tolman:1939jz,Oppenheimer:1939ne}. Although there are basic differences between hadronic and quark matter, such as the fact that hadrons are close to being non-relativistic whereas quarks are mostly ultra-relativistic, this does not necessarily lead to a clear signature in the EoS because quarks are expected to be strongly interacting, and our lack of theoretical understanding of strongly interacting systems leaves open the possibility that the EoS of quark matter could be similar to that of very dense hadronic matter, despite the generic softening due to the larger number of  species. As a result, there may be very little difference between the mass-radius relation of a hadronic star and that of a hybrid star with a quark matter core \cite{Alford:2004pf}. However, as we discuss in detail below, if there were a sufficiently strong first-order phase transition at the densities relevant to neutron stars then this could have observable consequences and provide an indication of the presence of a new state of matter at high densities, such as quark matter.

Measurements of both mass and radius for the same stars would greatly help to discriminate among possible EoSs. However, radius measurements are hampered by factors such as complexities in the modeling of the atmospheric composition, uncertainties in distance, magnetic fields, and inter-stellar absorption \cite{Ozel:2016oaf}. It has been shown that allowing for strong first-order phase transitions in the EoS decreases the inferred radii from observations of eight quiescent low-mass X-ray binaries in globular clusters, and in this case, the radius is likely smaller than 12 km~\cite{Steiner:2017vmg}; the NICER mission is expected to eventually reduce these uncertainties to the 5-10\% level \cite{Ozel:2015ykl}. There are so far two restrictive observational constraints on the stiffness of the EoS (or equivalently, the speed of sound): EoSs that are too soft were eliminated by the discovery of massive pulsars~\cite{Demorest:2010bx,Antoniadis:2013pzd,Fonseca:2016tux}, whereas EoSs that are very stiff are inconsistent with the pre-merger gravitational wave signal from the binary neutron star merger GW170817, which disfavored large tidal deformation (and large radii)~\cite{LIGO:2017qsa,LIGO:2018exr}.

\begin{figure}[htb]
\includegraphics[width=0.6\hsize]{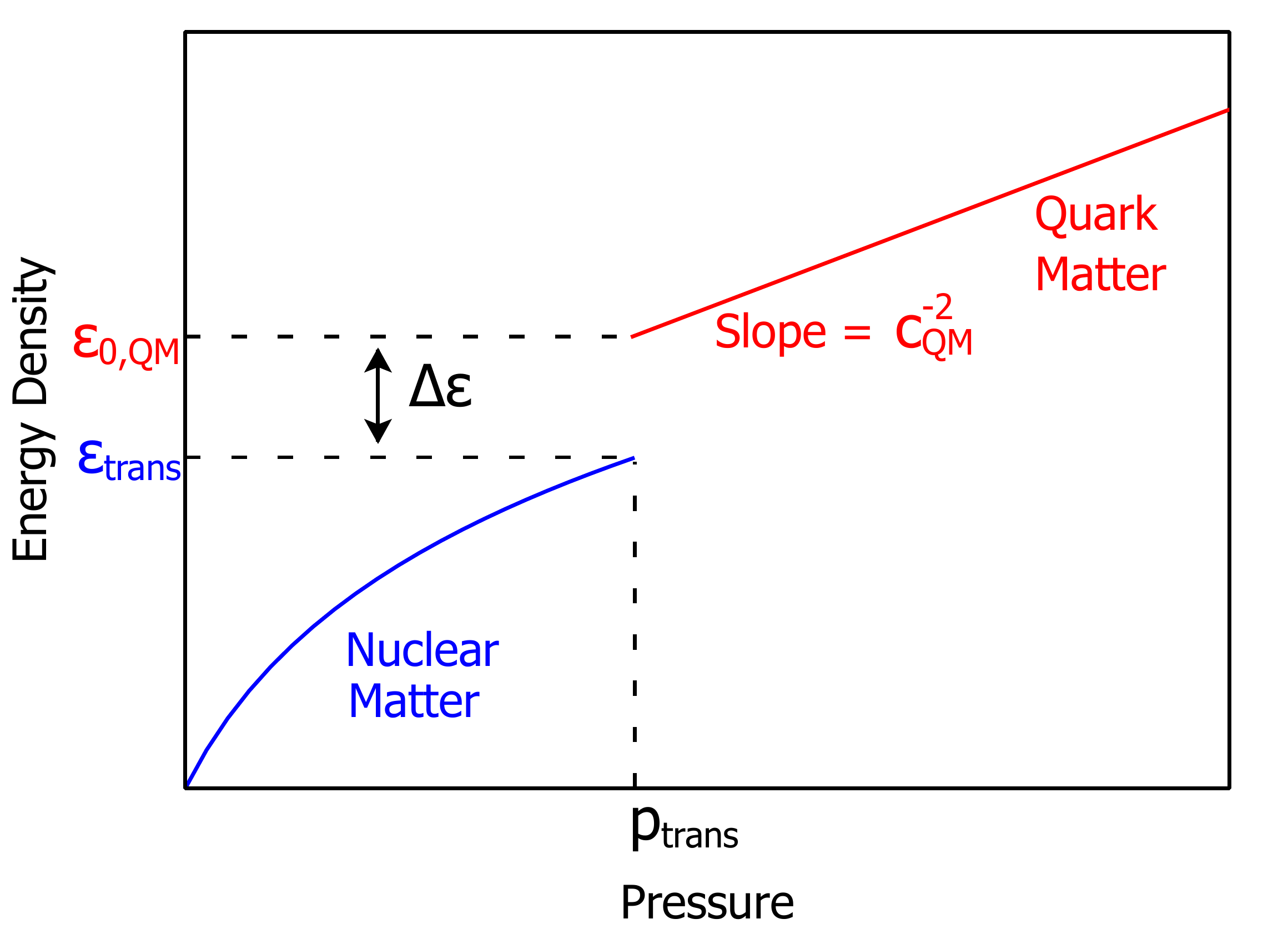}
\caption{(Color online) Equation of state $\ep(p)$ for dense matter with sharp first-order phase transition. In this parametrization the quark matter EoS is specified by the transition pressure $\ptrans$, the energy density discontinuity $\De\ep$, and the speed of sound in quark matter $\protect\cQM$ (assumed density-independent).}
\label{fig:eos-css}
\end{figure}

The tension between the existence of massive pulsars and the small tidal deformation inferred from the one observed merger (and possibly small radii from X-ray observations) can be reconciled by: \\
(i) a rapid switch from a soft EoS to a stiff EoS over a narrow pressure range in normal hadronic matter (the ``minimal scenario''), as indicated from e.g. computations of chiral effective field theory (EFT)~\cite{Hebeler:2010jx} and quantum Monte Carlo (QMC) methods~\cite{Gandolfi:2011xu}, or\\
(ii) a rapid crossover region of hadron-quark duality (the ``crossover scenario'') in e.g. quarkyonic matter or interpolated EoS~\cite{Masuda:2012ed,Fukushima:2015bda,Baym:2017whm,McLerran:2018hbz}, which typically yields simple mass-radius topology, or\\ (iii) strong first-order phase transition(s) at some supranuclear densities, creating an intrinsic softening of the EoS due to the finite energy density discontinuity (Fig.~\ref{fig:eos-css}) with sufficiently stiff quark matter above the transition.

Below we shall elaborate on this ``sharp phase-transition'' scenario. If the surface tension at the hadron-quark phase interface were low enough, such a transition would be smoothed out by the appearance of charge-separated mixed phases (Gibbs construction) \cite{Alford:2001zr,Maruyama:2007ss,Endo:2011em,Sotani:2010mx}. 
However, given the uncertainty in estimates of the surface tension \cite{Alford:2001zr,Mintz:2009ay,Lugones:2013ema,Ke:2013wga,Fraga:2018cvr} we will assume a sharp interface (Maxwell construction).

There have been many recent studies of the manifestations of a first-order phase transition, e.g., Refs.~\cite{Bandyopadhyay:2017dvi,Nandi:2017rhy,Paschalidis:2017qmb,Sen:2018yyq,Li:2018ayl,Alvarez-Castillo:2018pve,Wei:2018mxy,Ibanez:2018myp}. 
To conduct a model-independent survey of the observable signatures of such a transition, we will combine a plausible nuclear matter EoS at low densities with a generic parameterization of a first-order transition to a higher density phase. One example of this is the ``constant-speed-of-sound'' (CSS) parameterization (see Fig.~\ref{fig:eos-css}) \cite{Alford:2013aca} which contains only three parameters, specifying the critical pressure at which the transition occurs, the strength of the transition, and the ``stiffness'' of the high-density phase, where the speed of sound (which characterizes how rapidly pressure is rising with energy density) is assumed to be independent of density.
The CSS parametrization can be used as a generic language for relating different models to each other, and for expressing experimental and observational constraints. For hadronic matter we use the Dirac-Brueckner-Hartree-Fock (DBHF) EoS~\cite{GrossBoelting:1998jg} that satisfies the unitary gas limit~\cite{Kolomeitsev:2016sjl} and empirical constraints from low-density nuclear experiments.

First-order phase transition(s) can produce qualitative and quantitative features in the mass-radius curve for neutron stars. Qualitatively, a sharp transition to quark matter leads to a branch of hybrid stars (i.e. a quark matter core where the pressure exceeds $\ptrans$) \cite{Haensel:1983,Lindblom:1998dp}.
Depending on the energy density discontinuity and the speed of sound, there may or may not be a stable branch of hybrid stars, which may or may not be disconnected from the hadronic branch \cite{Alford:2013aca}.

\begin{figure*}[htb]
\parbox{0.5\hsize}{
\includegraphics[width=\hsize]{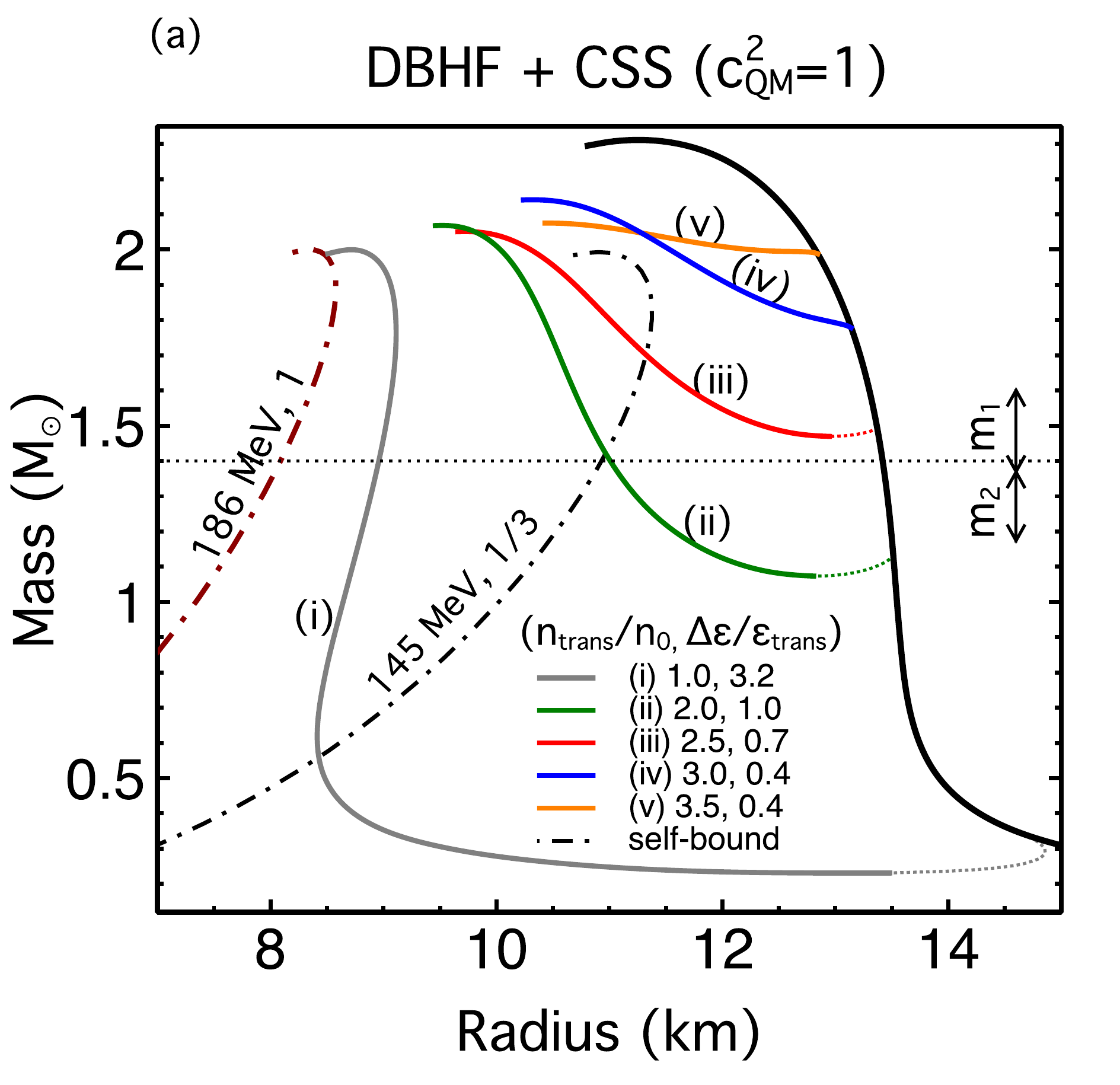}\\
}\parbox{0.52\hsize}{
\includegraphics[width=\hsize]{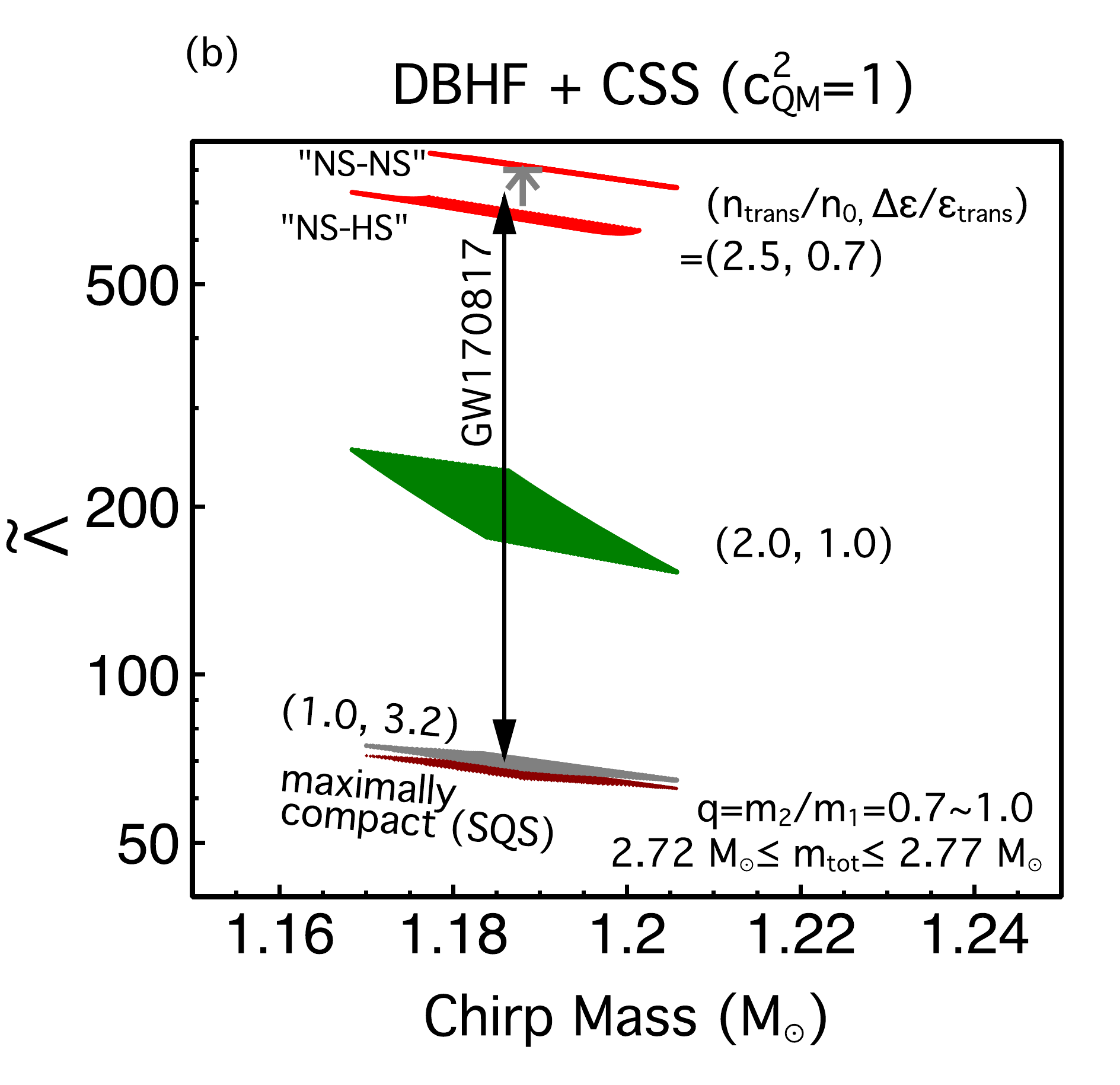}\\
}
\caption{(Color online) Panel (a): mass-radius relation for purely hadronic stars, strange quark stars, and hybrid stars. The DBHF EoS is used for the hadronic matter, and high-density quark phase is parameterized in the CSS framework with maximal speed of sound ($\cQMsq=1$). Ranges of the primary and secondary mass $(m_1, m_2)$ in GW170817 for low-spin priors are shown for comparison \cite{LIGO:2018wiz}. Panel (b): observables relevant to mergers: the weighted average dimensionless tidal deformability $\tilde{\La}$ as a function of the chirp mass $\Mchirp$ for each EoS employed in panel (a). Each colored quadrilateral covers the range of mass ratio $q$ and total mass $m_{\rm tot}$ found in the binary system of GW170817; the vertical line with double arrowheads  shows the 90\% confidence interval for $\tilde\La$ from the gravitational wave observation
of GW170817 \cite{LIGO:2018wiz}. See text for details.} 
\label{fig:mr-Lam-dbhf-css}
\end{figure*}

If the phase transition is sufficiently strongly first-order (large enough $\De \ep$) and occurs at low enough pressure then there will be a  disconnected hybrid branch, or ``third family'' of stars with smaller radii (and hence smaller tidal deformability), separated from the hadronic branch by a gap in $R$ where one should not detect any stable neutron stars (e.g., curves (i), (ii), (iii) in Fig.~\ref{fig:mr-Lam-dbhf-css}(a)). If there is a second phase transition then there might be a fourth-family of hybrid stars with even smaller radius and tidal deformability~\cite{Alford:2017qgh,Han:2018mtj}.

Mass and radius observations also lead to quantitative constraints on the CSS parameterization of the quark matter EoS. This has been explored in Ref.~\cite{Alford:2015gna}. Just requiring $\Mmax\geqslant 2\,\Msolar$ already rules out strongly first-order phase transitions at intermediate densities (see e.g. Fig.~5 of Ref.~\cite{Alford:2015gna}). If we assume that the speed of sound is always below the free-quark value $c_s^2=1/3$ (as QCD perturbation theory would suggest) then first-order transitions of any kind are possible only for very stiff hadronic equations of state, leading to speculation that $c_s^2$ must rise above $1/3$ in some density range \cite{Chamel:2012ea,Alford:2013aca,Bedaque:2014sqa,Alford:2015gna,Tews:2018kmu}. If a pulsar heavier than $2\,\Msolar$ were to be detected, that would yield even stronger constraints on possible first-order transitions, and if the form of the nuclear matter EoS were better known then measurements of the $M(R)$ relation of neutron stars would lead to correspondingly stronger constraints on the values of quark matter EoS parameters.

\section{Thermal evolution}
\label{sec:thermal}

\subsection{Predictions for different forms of matter}
In both quark and nuclear matter, across a broad range of temperatures, the dominant cooling mechanism at sufficiently high temperatures is neutrino emission from the bulk \cite{Yakovlev:2004iq}. The available cooling mechanisms can be classified into
\begin{itemize}
    \item fast cooling (neutrino emissivity $\eps \sim T^6$) due to direct Urca reactions \cite{Iwamoto:1980eb,Lattimer:1991ib},
    \item intermediate cooling ($\eps \sim T^7$), e.g. due to pair breaking,
    \item slow cooling ($\eps \sim T^8$ due to modified Urca reactions \cite{Friman:1978zq}, or higher powers of $T$).
\end{itemize}

In hadronic matter, the dominant neutrino emission mechanism at moderate density is modified Urca (slow cooling). Fast processes are generally suppressed because nucleons are fairly non-relativistic, making it difficult to conserve both energy and momentum in a direct Urca reaction. Above a threshold density which depends sensitively on the details of the nucleon interactions, fast cooling via direct Urca becomes possible.

Quark matter comes in different forms with differing properties. Phases with ungapped quarks have a high heat capacity $\sim \mu^2 T$ (from all the phase space at the quark Fermi surface) and show fast cooling via direct Urca, since the quarks are ultra-relativistic which facilitates momentum and energy conservation in direct Urca. In contrast, the color-flavor-locked (CFL) phase of quark matter is thermally inert: all the quarks are gapped, so there is a low heat capacity $\sim T^3$ arising from the superfluid phonons, and a low neutrino emissivity $\sim T^{15}$ \cite{Alford:2007xm}.

\subsection{Constraints from observations}

The cooling of young sources can be well described by a minimal neutron star model \cite{Page:2004fy} that includes intermediate cooling due to pair breaking emission from proton and/or neutron condensates \cite{Page:2009fu}. The thermal data from young sources is inconsistent with fast cooling. This sets tight constraints on the presence of quark matter phases with ungapped quarks in these sources.

In old sources in LMXBs the heat deposited onto the star's surface is generally immediately radiated off from the surface and cannot reach the star's interior, because the thermal conductivity in the outer crust drops dramatically with density close to the surface, effectively acting as a ``heat blanket''. However, light elements accreted on the surface catalyze as they are buried by subsequent accretion. Deep in the crust they transform by pycnonuclear reactions and the resulting energy release should heat the core of the star. Most LMXBs show outburst-quiescence cycles. Within this deep crustal heating scenario, it is assumed that over many cycles a steady state develops where deep crustal heating due to the accretion during outburst is balanced by the bulk neutrino cooling during quiescence \cite{Haensel90,Brown:1998qv,Yakovlev:2002ti}. The interesting result from these studies is that, while most sources are consistent with slow or intermediate cooling, there are ordinary soft X-ray transients, like 1H 1905+000 and SAX J1808.4--3658 that are so cold in quiescence that they require fast cooling \cite{Heinke10,Wijnands:2017jsc}. In quasi-persistent transients that undergo much longer outbursts driving the crust out of thermal equilibrium with the core, long-term monitoring can probe the core neutrino luminosity and, in some cases provide direct evidence for fast cooling as has been shown for the transient system MXB 1659-29 ~\cite{Brown:2017gxd}. In these systems, observations of the late time cooling of accreting neutron stars back to equilibrium can also constrain the heat capacity of the core, and hence the presence of very low heat-capacity phases such as CFL quark matter \cite{Cumming:2016weq}.

Forms of quark matter with ungapped quarks would naturally provide a fast cooling mechanism via direct Urca processes \cite{Yakovlev:2004iq}, but also in hadronic matter the Urca channel opens at sufficiently high density and, since the involved particles are close to being non-relativistic, could provide rates that are even an order of magnitude above those of quark Urca \cite{Lattimer:1991ib}. This could be too much to explain the cooling of these sources and precise future measurements might be able to distinguish these different options. 

Strange quark stars, which would be realized if the strange matter hypothesis holds \cite{Witten:1984rs}, could be ``bare'', meaning that their surface is a sharp transition from quark matter to vacuum without a low density crust. Such a scenario can probably be ruled out as a candidate for all observed thermal sources since the cooling would be extremely efficient. However, it is impossible to exclude that these sources exist in large numbers, since they would quickly be too cold to be observable. If strange stars have a hadronic crust elevated by electrostatic forces, their thermal properties should be virtually indistinguishable from hadronic or hybrid stars. 

Another possibility is a strangelet crust, composed of strangelets (quark nuggets) immersed in a sea of electrons \cite{Jaikumar:2005ne,Alford:2006bx,Alford:2008ge}. It seems likely that the cooling of such a crust will be similar to that of an ordinary nuclear  crust.

\section{Oscillations and spin-down}
\label{sec:oscillations}

Continuous gravitational waves from unstable oscillation modes, e.g.~r-modes \cite{Andersson:1997xt}, are very promising, since these modes are predicted to arise spontaneously in fast spinning compact sources unless there are strong dissipation mechanisms that can damp them. As well as emitting gravitational waves they would strongly affect the thermal and spin-down evolution of compact sources. Therefore multi-messenger observation of fast-spinning compact stars can indirectly probe the damping of these modes, which in turn is sensitively dependent on the phases of matter present inside them. 

\begin{figure}
\includegraphics*[width=7.8cm]{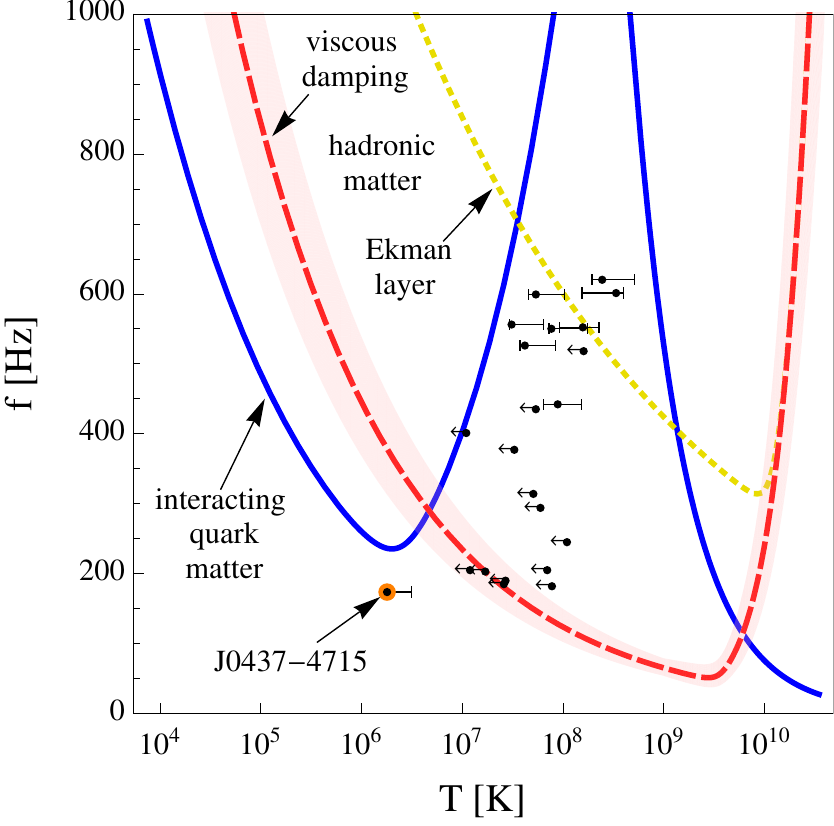}
\includegraphics*[width=7.8cm]{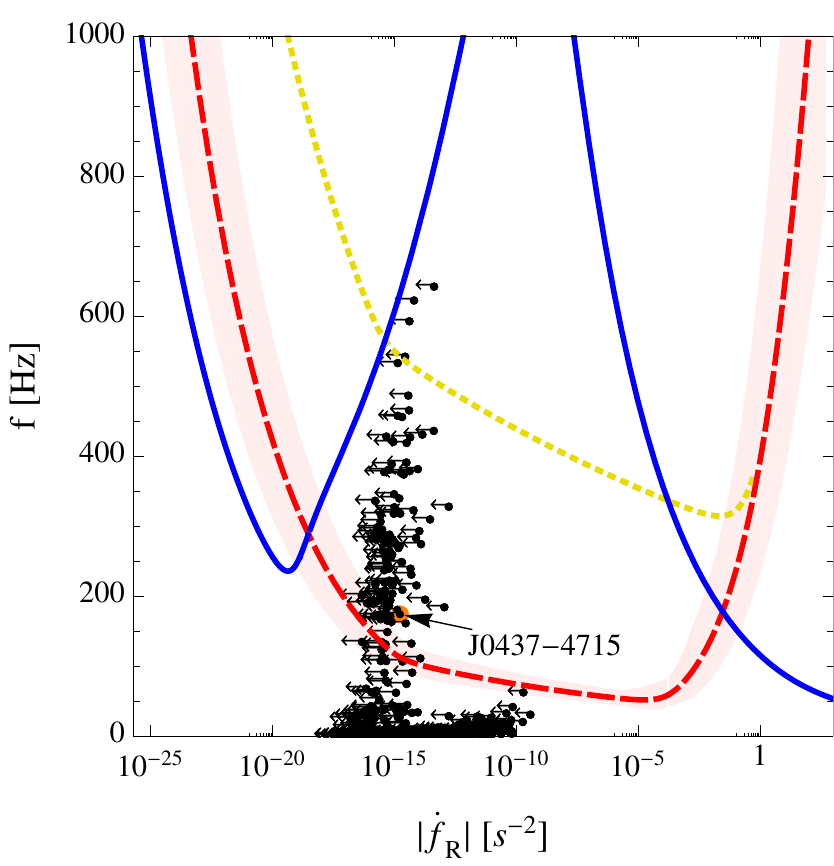}
\caption{
Boundaries of the r-mode instability
regions for different star compositions compared to pulsar data. \emph{Left:}
Standard static instability boundary compared to X-ray data \cite{Haskell:2012,Tomsick:2004pf}
with error estimates from different envelope models \cite{Gudmundsson:1983ApJ,Potekhin:1997}.
\emph{Right:} Dynamic instability boundary in timing parameter space
compared to radio data \cite{Manchester:2004bp} (all data points
are upper limits for the r-mode component of the spin-down). The curves
represent: $1.4\,\Msolar$ neutron star (NS) with standard viscous
damping \cite{Shternin:2008es,Sawyer:1989dp} (dashed) and with additional
boundary layer rubbing \cite{Lindblom:2000gu} at a rigid crust (dotted)
as well as $1.4\,\Msolar$ strange star (SS) with long-ranged non-Fermi liquid (NFL)
interactions causing enhanced damping \cite{Heiselberg:1993cr,Schwenzer:2012ga}
(using $\al_{s}\!=\!1$) (solid)---more massive stars are not qualitatively
different. The thin curves show for the standard neutron star exemplarily
the analytic approximation for the individual segments. The encircled
points denote the only millisecond radio pulsar J0437-4715
with a temperature estimate.}
\label{fig:instability-regions}
\end{figure}

An important result, known since the discovery of the r-mode instability, is that the dissipation in minimal models of neutron stars is not sufficient to damp these modes at typical temperatures of observed millisecond sources $T_s\sim 10^5-10^6\,{\rm K}$ \cite{Lindblom:1998wf}. This is illustrated, along with pulsar timing data that leads to a similar conclusion \cite{Alford:2013pma}, in Fig.~\ref{fig:instability-regions} (taken from \cite{Alford:2014nba}). Stars with spin frequency and temperature or spindown rate that places them inside the V-shaped region for a given EoS would experience the r-mode instability. There are two ways to account for the observation of stars within this region:
(1) there are additional damping mechanisms that shift the instability curves upwards; (2) although small-amplitude r-modes are unstable inside the V-shaped regions, non-linear mechanisms saturate their amplitude at a low value $\al_{\rm sat}$. 

Spectral X-ray observations of LMXBs and millisecond pulsars have put a bound on the saturation amplitude $\al_{\rm sat} \lesssim 10^{-9}-10^{-8}$  \cite{Mahmoodifar:2013quw,Schwenzer:2016tkf,Bhattacharya:2017nxc}. This bound is obtained by looking for the heating that would be produced by the  dissipation required to saturate the mode. Many mechanisms have been proposed to saturate the r-modes. Examples include the non-linear coupling of the r-mode to daughter modes \cite{Arras:2002dw,Bondarescu:2013xwa} that are viscously damped, or the rubbing of the fluid in the core along the solid crust \cite{Lindblom:2000gu}. However, these proposed mechanisms lead to saturation amplitudes $\al_{\rm sat} \gtrsim 10^{-6}$, which is several orders of magnitude away from the values required to explain the data. 
This striking discrepancy provides probably the clearest sign for new physics in the neutron star interior to date.  

Quark matter is a prime candidate for the required enhanced damping: we see in Fig.~\ref{fig:instability-regions} that the observed stars are all outside the instability region for interacting quark matter. The relevant damping mechanism is bulk viscosity due to non-leptonic weak interactions, which has a resonant peak at temperatures in the $10^7$ to $10^9$\,K range. 
Another way that quark matter can account for the data is by imposing a very low saturation amplitude on the r-modes. The relevant damping mechanism is dissipation from the periodic conversion of hadronic into quark matter in hybrid stars \cite{Alford:2014jha}, which provides the strongest known saturation mechanism and can easily provide saturation amplitudes consistent with the X-ray bounds mentioned above.

The very low bounds on the r-mode amplitudes in old millisecond pulsars make a direct gravitational wave detection of these sources unlikely with the current generation of detectors \cite{Alford:2014pxa}. However, young sources are much more promising targets for direct searches. It has been shown that r-modes in standard neutron stars can provide a quantitative explanation for the low spin frequencies of observed young pulsars \cite{Alford:2012yn}. There are several young sources that, if they were spinning down via r-modes, would emit gravitational waves that could be seen by current detectors and therefore provide promising targets for dedicated searches \cite{Alford:2012yn}.

Just as already indicated by purely thermal observables, also the spin data suggest that there may be different classes of sources. According to this picture, young sources would be born light, as suggested by the canonical masses $\lesssim 1.4\,\Msolar$ in double neutron star binaries \cite{Ozel:2016oaf}, and therefore initially would have a purely hadronic composition, guaranteeing a low neutrino emissivity and weak dissipation \cite{Page:2004fy,Alford:2012yn}. However, if their mass increases sufficiently by recycling in mixed binaries \cite{Ozel:2016oaf}, they would develop a quark matter core or completely convert to a strange star \cite{Witten:1984rs}, with strong damping of r-modes and fast cooling. Such a scenario could naturally explain the presently compiled astrophysical data \cite{Haskell:2012,Manchester:2004bp}.

Another interesting observable is the maximum spin frequency of observed recycled millisecond pulsars. While based on their structural stability neutron stars could theoretically spin with Kepler frequencies significantly above a kHz, the fastest observed source PSR J1748-2446ad spins with a frequency of 716 Hz and there are several sources close to this value. This suggests that another mechanism imposes a spin limit \cite{Bildsten:1998ey}. R-modes can provide such a mechanism. In particular, if quark matter is present in the source, the maximum of the stability window due to non-leptonic bulk viscosity \cite{Alford:2010fd} could provide a natural explanation for the observational data \cite{Andersson:2001ev}. In contrast to other r-mode observables which prove to be strikingly insensitive to unknown physics \cite{Alford:2012yn}, this maximum is unfortunately very sensitive to poorly constrained microphysical details \cite{Alford:2010fd} so that a quantitative comparison is complicated at present.

High initial spin frequencies can also drive large magnetic fields. While external dipole fields of magnetars, as inferred from their spin-down rate, can reach values O($10^{15}$ G), the internal fields could be even larger. In case fields O($10^{17}-10^{18}$ G) would be reached in the interior~\cite{Lai:1991cold,Bocquet:1995je,Cardall:2000bs,Ferrer:2010wz}, the magnetic fields could affect the phase structure of dense matter. However, the fact that known magnetars are sufficiently young, would require that neutron stars are already born with sufficiently large masses to support quark matter in their cores, which would only be consistent with thermal data for completely gapped phases as the important CFL phase \cite{Alford:1998mk}. For the latter it has been shown that the magnetic field can reduce the symmetry of the pairing and can at even larger field strength even lead to interesting phases with extended gluonic field configurations \cite{Ferrer:2012wa}. Characteristic properties of these phases might in the future provide signatures for quark matter.

\section{Solid phases}

A compact star with a solid crust can sustain deformations (``mountains''), which could arise due to local accretion and/or strong magnetic fields. The size of deformations is typically parametrized by an ellipticity parameter $\eps$, whose upper limit is set by the breaking strain of the crust. In ordinary neutron stars the shear modulus and breaking strain are relatively low so that only a moderate ellipticity $\eps \lesssim 10^{-5}$ can be sustained against gravity \cite{Horowitz:2009ya,JohnsonMcDaniel:2012wg}. However, one of the leading candidates for quark matter at moderate density is the crystalline color superconducting phase \cite{Rajagopal:2006ig,Anglani:2013gfu}, where the color-superconducting condensate varies periodically in space, forming a crystal that is much more rigid, having a thousand times larger shear modulus \cite{Mannarelli:2007bs}. The breaking strain  is still very uncertain, but the lack of observed gravitational radiation from known sources indicates that they do not contain maximally strained crystalline cores \cite{Haskell:2007sh,Knippel:2009st}.

One type of observational constraint is the extraction of upper limits on the ellipticity of known pulsars, obtained by looking for the gravitational waves that would have been emitted if they had non-axisymmetric deformations. These provide restrictive bounds $\eps \lesssim 2\times 10^{-7}$ on the ellipticity of several sources  \cite{Abbott:2018bwn}. There is therefore no indication that these sources contain crystalline quark matter, and these limits could provide evidence of its absence if one could show that the formation process should lead to larger deformations of the crystalline component. It has been shown in Ref.~\cite{Haskell:2017ajb} that the increase in the spin-down rate of PSR J1023+0038 during its LMXB phase is compatible with gravitational wave emission, either due to the creation of a ``mountain'' during the accretion phase or an r-mode of amplitude $\al\approx 5\times 10^{-8}$.

Another possible manifestation of the presence of crystalline quark matter is the occasional ``glitches'' observed in spinning neutron stars, where the star's spin rate suddenly increases (for a review see Ref.~\cite{Haskell:2015jra}). The standard explanations for this involve exchange of angular momentum between the solid nuclear crust and the neutron superfluid, which coexist in the inner crust. However, it has also been proposed \cite{Mannarelli:2007bs} that a similar mechanism could operate in a hybrid star, where a crystalline quark matter core could coexist with the color-flavor-locked superfluid phase of quark matter.

\section{Neutron star mergers}
\label{sec:mergers}

Neutron star mergers \cite{Baiotti:2016qnr} are the only processes that provide direct gravitational wave data at present \cite{LIGO:2017qsa}. So far only the inspiral phase has been observed. The inspiral phase of a binary neutron-star system produces gravitational waves that carry information about the tidal deformability of the stars, which is determined by the equation of state and strongly correlated with their radii \cite{LIGO:2017qsa,LIGO:2018exr,LIGO:2018wiz}.

Panel (b) of Fig.~\ref{fig:mr-Lam-dbhf-css} illustrates how the tidal deformability, measured by the ``combined tidal deformability'' $\tilde{\La}=\frac{16}{13}\left[(m_1+12m_2)m_1^4\,\La_1+(m_2+12m_1)m_2^4\,\La_2\right]/{(m_1+m_2)^5}$,
relates to the CSS parameters of a quark matter EoS.   The colored quadrilaterals show the range of tidal deformabilities for different choices of the masses of the stars in the binary for each of the equations of state whose mass-radius relations are shown in panel (a). 
For each equation of state we scan all possible combinations of the primary mass $m_1 \in [1.36, 1.60]\, \Msolar$ and secondary mass $m_2 \in [1.16, 1.36]\, \Msolar$, imposing bounds on the total mass $m_{\rm tot}=2.73_{-0.01}^{+0.04}\, \Msolar$ and mass ratio $q\in[0.7,1.0]$ for low-spin priors \cite{LIGO:2018wiz}, and compute $\La_1$, $\La_2$ and eventually $\tilde{\La}$.
There is no blue quadrilateral because the mass range for GW170817 never reaches the beginning of the hybrid branch for EoS (iv). 
The double-headed arrow shows the symmetric 90\% credible interval $\tilde\La (\Mchirp=1.186_{-0.001}^{+0.001}\Msolar)=300_{-230}^{+420}$ in the low-spin case \cite{LIGO:2018wiz}, and the other arrow represents the upper limit for low-spin priors reported in the original detection paper \cite{LIGO:2017qsa}.

One interesting feature of the combined tidal deformability is that, in a context where the total mass of the binary system is well constrained, it is sensitive to the EoS of dense matter but generally insensitive to the poorly-known mass ratio $q=m_2/m_1$~\cite{Radice:2017lry, Raithel:2018ncd}. However, first-order transitions provide an exception, as indicated by the presence of two separate predictions (uppermost (red) quadrilaterals in  Fig.~\ref{fig:mr-Lam-dbhf-css}) that derive from the same EoS, (iii). 

The hybrid star is more compact, leading to a lower deformability. In this case the (NS-NS) range of $\tilde\La$ is outside the 90\% confidence interval from the data but the (NS-HS) case is within it. Thus a hadronic EoS that seems to be excluded by tidal deformability data can be ``rescued'' by a sharp transition to quark matter. If in the future more data from advanced LIGO of multiple BNS mergers were to provide a refined range estimate of $\tilde{\La}$ for given chirp mass, we anticipate increasingly better constraints on the phase transition parameters and quark matter EoS. Recently tidal deformation of compact stars with strange quark matter has been studied in various models \cite{Lau:2017qtz, Annala:2017tqz,Paschalidis:2017qmb,Nandi:2017rhy,Alvarez-Castillo:2018pve,Burgio:2018yix,Gomes:2018eiv,Lau:2018mae}. Refs. \cite{Most:2018hfd,Tews:2018iwm,Sieniawska:2018zzj,Zhao:2018nyf,Christian:2018jyd,Montana:2018bkb} utilized phenomenological parameterization similar to CSS for a first-order phase transition, leading to consistent results with what we present here.

In the slow-rotation approximation, the dimensionless tidal deformability $\La$ and the dimensionless moment of inertia $\bar{I}= I/M^3$ of neutron stars without phase transitions are related by EoS-independent universal relation (``I-Love'' relation) to within $1\%$~\cite{Yagi:2013awa,Yagi:2016bkt}. However, in the presence of a strong first-order phase transition, the deviation from the ``I-Love'' universal relation can be as large as $~2\%$ for a single transition and $~9\%$ for sequential transitions~\cite{Paschalidis:2017qmb,Wei:2018dyy,Han:2018mtj}. Therefore, possible precise measurements of moments of inertia via pulsar timing, in particular for PSR J0737--3039A with known mass $M_{A}=1.338\, \Msolar$ and low spin frequency $f=44 \,\rm{Hz}$ in a double pulsar system~\cite{Lattimer:2004nj,Watts:2014tja}, which combined with testing the universal relation, hold promise to distinguish among EoS models with and without strong phase transitions. Novel solid phases, e.g.~crystalline color-superconducting quark matter~\cite{Alford:2000ze,Casalbuoni:2003wh}, may also alter the tidal properties affecting the universal relation~\cite{Lau:2017qtz,Lau:2018mae}; it is also found that strong first-order transition induces deviation from the ``I-Q'' universal relation (where $Q$ is the quadrupole moment) for fast rotating neutron stars~\cite{Bandyopadhyay:2017dvi}. The dimensionless spin parameter $j=J/M^2$ has an upper bound $\sim 0.7$ for normal hadronic stars, but self-bound quark stars allow much larger values exceeding 1~\cite{Lo:2010bj,Doneva:2013rha}, which might be important for e.g. interpreting constraints from mergers \cite{LIGO:2017qsa,LIGO:2018wiz} or maximum mass of neutron stars \cite{Bozzola:2019tit}. 

The late stage of the inspiral, where tidal effects become large, could in principle also probe dynamic observables. Tidal effects can for instance resonantly excite oscillation modes in the inspiraling sources \cite{Kokkotas:1995xe}, and in particular the fundamental mode (f-mode) can have a significant overlap with the tidal deformations \cite{Chirenti:2016xys} and likewise has frequencies in the kHz regime \cite{Kokkotas:1999bd}. Moreover, there could be interesting observable effects due to the shattering of a solid crust \cite{Tsang:2011ad}.

The actual merger phase has not yet been observed. Depending on the total mass of the system, a meta-stable merger product may be formed, stabilized by angular momentum and/or thermal pressure \cite{Baiotti:2016qnr}. The gravitational wave signal from this phase is expected to be much more complicated and therefore harder to target in gravitational wave searches, which become much harder if one does not have detailed knowledge of the expected signal. However, hydrodynamic simulations show characteristic peaks in the power spectrum \cite{Baiotti:2016qnr}, so that a future detection seems possible.

Model-dependent simulations with first-order hadron/quark phase transition predict a unique evolution of temperature and density in the merger remnant, and different spectra in post-merger gravitational wave signals~\cite{Most:2018eaw,Bauswein:2018bma}. However, it is not clear at present if this can provide a clear signature for quark matter taking into account the significant uncertainties in the process.

Since the actual merger results in a highly excited merger product, with large entropy, velocity gradients and density compressions, dynamic material properties of the matter could be relevant in this phase. First estimates~\cite{Alford:2017rxf} show that in particular bulk viscosity could be very important for the evolution and locally dissipate a significant fraction of the kinetic energy on a dynamic timescale if the stars are made of hadronic matter with modified Urca interactions. The latter requirement is again in accordance with the low masses $\lesssim 1.4\,\Msolar$ needed for a formation of a meta-stable merger product instead of a prompt collapse of the system. In case of quark matter in turn the standard bulk viscosity would be orders of magnitude lower \cite{Alford:2010gw}, so that future gravitational wave signals could allow us to distinguish different compositional scenarios. Strangeness production in the massive merger product \cite{Alcock:1986hz} and phase-conversion dissipation due to density oscillations \cite{Alford:2014jha} could be even better probes for quark matter, since these processes are very efficient on millisecond dynamic merger time scales and could have a clearly observable impact on the merger dynamics and the resulting gravitational wave signal.

\section{Conclusions}
There is as yet no firm evidence for quark matter in neutron star cores. This is mainly because of the lack of direct probes of the opaque neutron star interior: the available messengers are all indirect. Another challenge is the lack of clear qualitative difference between hadronic and quark phases, but theoretical analysis predicts dramatic quantitative differences between the material properties of the various phases of dense matter, and this makes it possible to develop observational criteria for the presence of certain types of quark matter.

There are first promising astrophysical indications---both from the surprisingly low temperatures of some LMXBs and from the unexplained r-mode damping in all fast millisecond pulsars---that a minimal neutron star composition is insufficient. Quark matter could in both cases provide a consistent explanation for these observations. The steadily growing body of astrophysical data and supported laboratory experiments (e.g. the forthcoming PREX-II~\cite{Abrahamyan:2012gp,Fattoyev:2017jql}) should eventually allow us to narrow down the options by combining these various observations. In particular, the arrival of direct gravitational wave observations is expected to provide significant new constraints.

\section*{Acknowledgements} 
This research is supported by the U.S. Department of Energy, Office of Science, Office of Nuclear Physics under Award Number \#DE-FG02-05ER41375, the Turkish Research Council TUBITAK via the project number 117F312, National Science Foundation Grant PHY-1630782, and the Heising-Simons Foundation, Grant 2017-228.

\bibliographystyle{unsrt}

\bigskip
\bibliography{quark_matter}{}

\end{document}